%% file: egpaper_arxiv_after_final.tex
\documentclass[10pt,twocolumn,letterpaper]{article}
\pdfoutput=1
\usepackage{cvpr}
\usepackage{times}
\usepackage{epsfig}
\usepackage{graphicx}
\usepackage{amsmath}
\usepackage{amssymb}

\usepackage{array}
\usepackage{bbm}
\usepackage{enumitem}
\usepackage{multirow}
\usepackage{relsize}
\usepackage{subfloat}
\usepackage[
 font=small
 ]{subfig}
\usepackage[title]{appendix}
\newif\ifarxiv
\arxivtrue

\ifarxiv
\newcommand{\arxiv}[1]{#1}
\else
\newcommand{\arxiv}[1]{}
\fi
\DeclareMathOperator*{\argmax}{arg\,max}

\newcolumntype{P}[1]{>{\centering\arraybackslash}p{#1}}
\newcolumntype{M}[1]{ >{\centering\arraybackslash} m{#1} }

\newcommand{\hl}[1]{#1}
\usepackage[title]{appendix}

\usepackage[breaklinks=true,bookmarks=false]{hyperref}

\cvprfinalcopy 

\def\cvprPaperID{8704} 

\begin{document}

\title{Painting Many Pasts: Synthesizing Time Lapse Videos of Paintings}

\author{
Amy Zhao\\
MIT\\
{\tt\small xamyzhao@mit.edu}
\and
\hspace{40pt}
Guha Balakrishnan\\
\hspace{40pt}
MIT\\
\hspace{40pt}
{\tt\small balakg@mit.edu}
\and
\hspace{40pt}
Kathleen M. Lewis\\
\hspace{40pt}
MIT\\
\hspace{40pt}
{\tt\small kmlewis@mit.edu}
\and
Fr\'edo Durand\\
MIT\\
{\tt\small fredo@mit.edu}
\and
John V. Guttag\\
MIT\\
{\tt\small guttag@mit.edu}
\and
Adrian V. Dalca\\
MIT, MGH\\
{\tt\small adalca@mit.edu}
}
\maketitle

\begin{abstract}
We introduce a new video synthesis task: synthesizing time lapse videos depicting how a given painting might have been created. Artists paint using unique combinations of brushes, strokes, and colors. There are often many possible ways to create a given painting. Our goal is to learn to capture this rich range of possibilities. 

Creating distributions of long-term videos is a challenge for learning-based video synthesis methods. We present a probabilistic model that, given a single image of a completed painting, recurrently synthesizes steps of the painting process. We implement this model as a convolutional neural network, and introduce a novel training scheme to enable learning from a limited dataset of painting time lapses. We demonstrate that this model can be used to sample many time steps, enabling long-term stochastic video synthesis. We evaluate our method on digital and watercolor paintings collected from video websites, and show that human raters find our synthetic videos to be similar to time lapse videos produced by real artists. Our code is available at \texttt{https://xamyzhao.github.io/timecraft}.
\end{abstract}

\input{intro}
\input{related_work}
\input{problem_definition}

\input{method}
\input{experiments}
\input{discussion}
{\small
\bibliographystyle{ieee_fullname}
\bibliography{egbib}
}
\clearpage
\appendix
\begin{appendices}
\input{supplementary_derivation}
\input{supplementary_details}
\input{supplementary_qualitative}

\end{appendices}

\end{document}

%% file: intro.tex
\section{Introduction}
Skilled artists can often look at a piece of artwork and determine how to recreate it. In this work, we explore whether we can use machine learning and computer vision to mimic this ability. We define a new video synthesis problem: \textit{given a painting, can we synthesize a time lapse video depicting how an artist might have painted it?}

\begin{figure}
\centering
\hspace{-5pt}\includegraphics[width=1\linewidth]{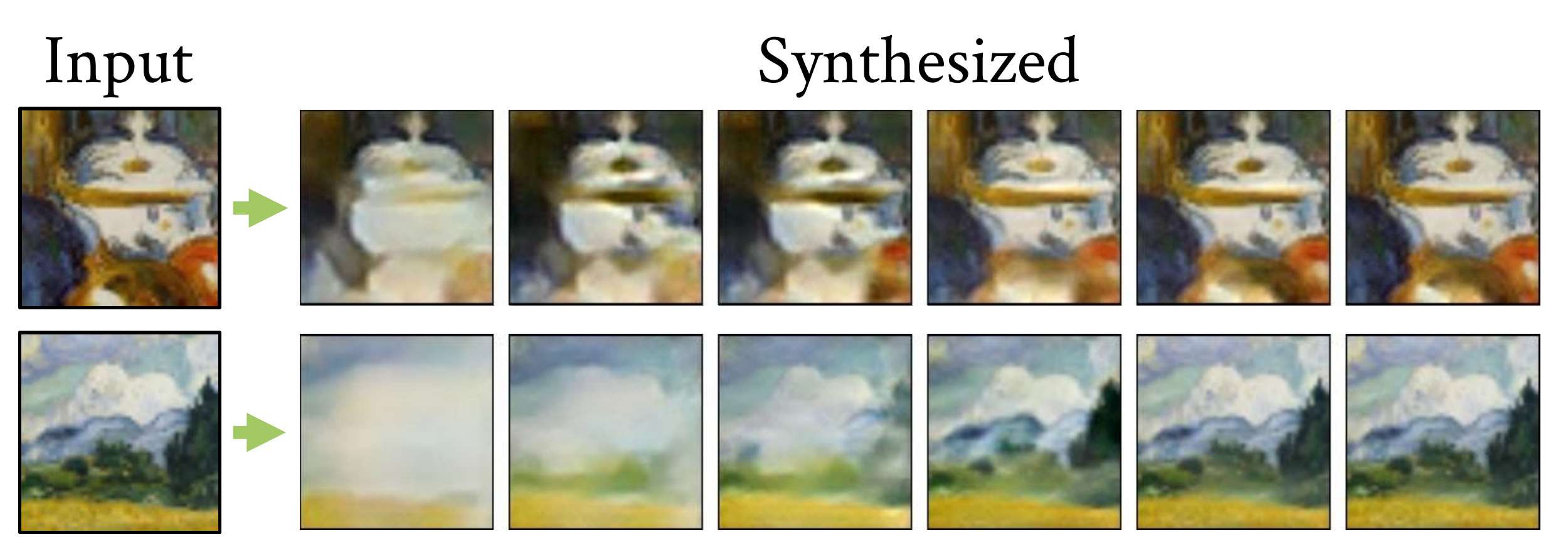}\vspace{-5pt}
\caption{We present a probabilistic model for synthesizing time lapse videos of paintings. We demonstrate our model on \textit{Still Life with a Watermelon and Pomegranates} by Paul Cezanne (top), and \textit{Wheat Field with Cypresses} by Vincent van Gogh (bottom).\vspace{-16pt}}\label{fig:chap4_teaser}
\end{figure}
Artistic time lapses present many challenges for video synthesis methods. There is a great deal of variation in how people create art. Suppose two artists are asked to paint the same landscape. One artist might start with the sky, while the other might start with the mountains in the distance. One might finish each object before moving onto the next, while the other might work a little at a time on each object. During the painting process, there are often few visual cues indicating where the artist will apply the next stroke. The painting process is also long, often spanning hundreds of paint strokes and dozens of minutes. 

In this work, we present a solution to the painting time lapse synthesis problem. We begin by defining the problem and describing its unique challenges. We then derive a principled, learning-based model to capture a distribution of steps that a human might use to create a given painting. We introduce a training scheme that encourages the method to produce realistic changes over many time steps. We demonstrate that our model can learn to solve this task, even when trained using a small, noisy dataset of painting time lapses collected from the web. We show that human evaluators almost always prefer our method to an existing video synthesis baseline, and often find our results indistinguishable from time lapses produced by real artists.

This work presents several technical contributions:
\begin{enumerate}[leftmargin=15pt,itemsep=1pt]
\item We \hl{use a} probabilistic model to capture stochastic decisions made by artists, thereby capturing a distribution of plausible ways to create a painting.
\item Unlike work in future frame prediction or frame interpolation, we synthesize long-term videos spanning dozens of time steps and many real-time minutes.
\item We demonstrate a model that successfully learns from painting time lapses ``from the wild.'' This data is small and noisy, having been collected from uncontrolled environments with variable lighting, spatial resolution and video capture rates.
\end{enumerate}

%% file: related_work.tex
\section{Related work}
To the best of our knowledge, this is the first work that models and synthesizes distributions of videos of the past, given a single final frame. The most similar work to ours is a recent method called \textit{visual deprojection}~\cite{balakrishnan2019visual}. Given a single input image depicting a temporal aggregation of frames, their model captures a distribution of videos that could have produced that image. We compare our method to theirs in our experiments. Here, we review additional related research in three main areas: video prediction, video interpolation, and art synthesis.

\subsection{Video prediction}
Video prediction, or future frame prediction, is the problem of predicting the next frame or few frames of a video given a sequence of past frames. Early work in this area focused on predicting motion trajectories \cite{bennewitz2002learning,gaffney1999trajectory,liu2010sift,vasquez2004motion,walker2014patch} or synthesizing motions in small frames \cite{michalski2014modeling,mittelman2014structured,sutskever2009recurrent}. Recent methods train convolutional neural networks on large video datasets to synthesize videos of natural scenes and human actions \cite{liu2017video,mathieu2016deep,ranzato2014video,villegas2017learning,vondrick2016generating}. \hl{A recent work on time lapse synthesis focuses on outdoor scenes \cite{nam2019end}, simulating illumination changes over time while keeping the content of the scene constant. In contrast, creating painting time lapses requires adding content while keeping illumination constant. Another recent time lapse method} outputs only a few frames depicting specific physical processes: melting, rotting, or flowers blooming~\cite{zhou2016learning}. 

Our problem differs from video prediction in several key ways. First, most \hl{video} prediction methods focus on short time scales, synthesizing frames on the order of seconds into the future, and encompassing relatively small changes. In contrast, painting time lapses span minutes or even hours, and depict dramatic content changes over time. Second, most video predictors output a single most likely sequence, making them ill-suited for capturing a variety of different plausible painting trajectories. One study~\cite{xue2016visual} uses a conditional variational autoencoder to model a distribution of plausible future frames of moving humans. We build upon these ideas to model \hl{painting changes} across multiple time steps. Finally, \hl{video} prediction methods focus on natural videos, which depict of motions of people and objects~\cite{liu2017video,mathieu2016deep,ranzato2014video,villegas2017learning,vondrick2016generating,xue2016visual} or physical processes \cite{nam2019end,zhou2016learning}. The input frames often contain visual cues about how the motion, action or physical process will progress, limiting the space of possibilities that must be captured. In contrast, snapshots of paintings provide few visual cues, leading to many plausible \hl{future} trajectories.

\subsection{\hl{Video} frame interpolation}
Our problem can be thought of as a long-term frame interpolation task between a blank canvas and a completed work of art, with many possible painting trajectories between them. In \hl{video} frame interpolation, the goal is to temporally interpolate between two frames in time. Classical approaches focus on natural videos, and estimate dense flow fields~\cite{baker2011database,werlberger2011optical,yu2013multi} or phase \cite{meyer2015phase} to guide interpolation. More recent methods use convolutional neural networks to directly synthesize the interpolated frame \cite{niklaus2017video}, or combine flow fields with estimates of scene information~\cite{jiang2018super,niklaus2018context}. Most frame interpolation methods predict a single or a few intermediate frames, and are not easily extended to predicting long sequences, or predicting distributions of sequences. 

\subsection{Art synthesis}
The graphics community has long been interested in simulating physically realistic paint strokes in digital media. Many existing methods focus on physics-based models of fluids or brush bristles \cite{baxter2004viscous,baxter2004versatile,
chen2015wetbrush,chu2005moxi,way2001synthesis,xu2002solid}. More recent learning-based methods leverage datasets of real paint strokes \cite{kim2010example,lu2013realbrush,zheng2017example}, often posing the artistic stroke synthesis problem as a texture transfer or style transfer problem~\cite{ando2010segmental,lukavc2015brushables}. Several works focus on simulating watercolor-specific effects such as edge darkening \cite{montesdeoca2017edge,wang2014towards}. We focus on capturing large-scale, long-term painting processes, rather than fine-scale details of individual paint strokes. 

In style transfer, images are transformed to simulate a specific style, such as a painting-like style~\cite{hertzmann1998painterly,huang2015synthesis} or a cartoon-like style~\cite{zhang2017data}. More recently, neural networks have been used for generalized artistic style transfer~\cite{gatys2015neural,zhu2017unpaired}. We leverage insights from these methods to synthesize a realistic progressions of paintings.

\hl{Several recent papers apply reinforcement learning or similar techniques to the process of painting. \hl{These approaches involve designing} parameterized brush strokes, and then training an agent to apply strokes to produce a given painting \cite{ganin2018synthesizing,huang2019learning,LPaintB2019,PaintBot2019,ArtistAgent2012,strokestyle2015,zheng2018strokenet}.} Some works focus on specific artistic tasks such as hatching or other repetitive strokes~\cite{jodoin2002hatching,xing2014autocomplete}. These approaches require careful hand-engineering, and are not optimized to produce varied or realistic painting progressions. In contrast, we learn a broad set of effects from real painting time lapse data.

%% file: problem_definition.tex
\section{Problem overview}
\begin{figure}
\centering
\includegraphics[width=0.97\linewidth]{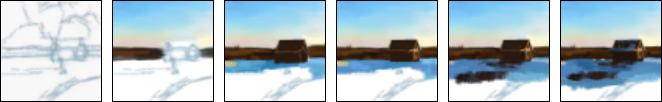}\vspace{2pt}
\includegraphics[width=0.97\linewidth]{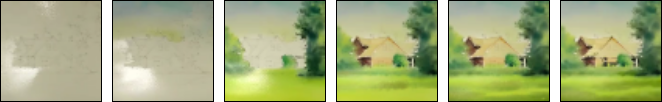}\vspace{2pt}
\includegraphics[width=0.97\linewidth]{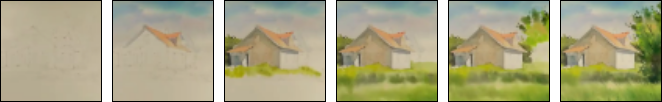}\vspace{3pt}
\includegraphics[width=0.97\linewidth]{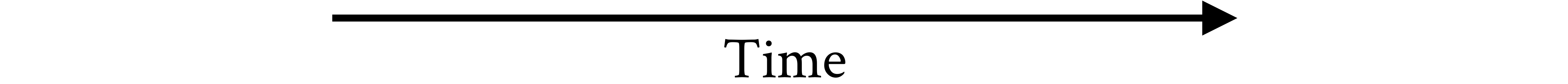}\vspace{-6pt}
\caption{\textbf{Several real painting progressions of similar-looking scenes}. Each artist fills in the house, sky and field in a different order.}\label{fig:problem_trajectories}
\end{figure}
Given a completed painting, our goal is to synthesize different ways that an artist might have created it. We work with recordings of digital and watercolor painting time lapses collected from video websites. Compared to natural videos of scenes and human actions, videos of paintings present unique challenges.

\begin{figure}
\centering
\includegraphics[width=\linewidth]{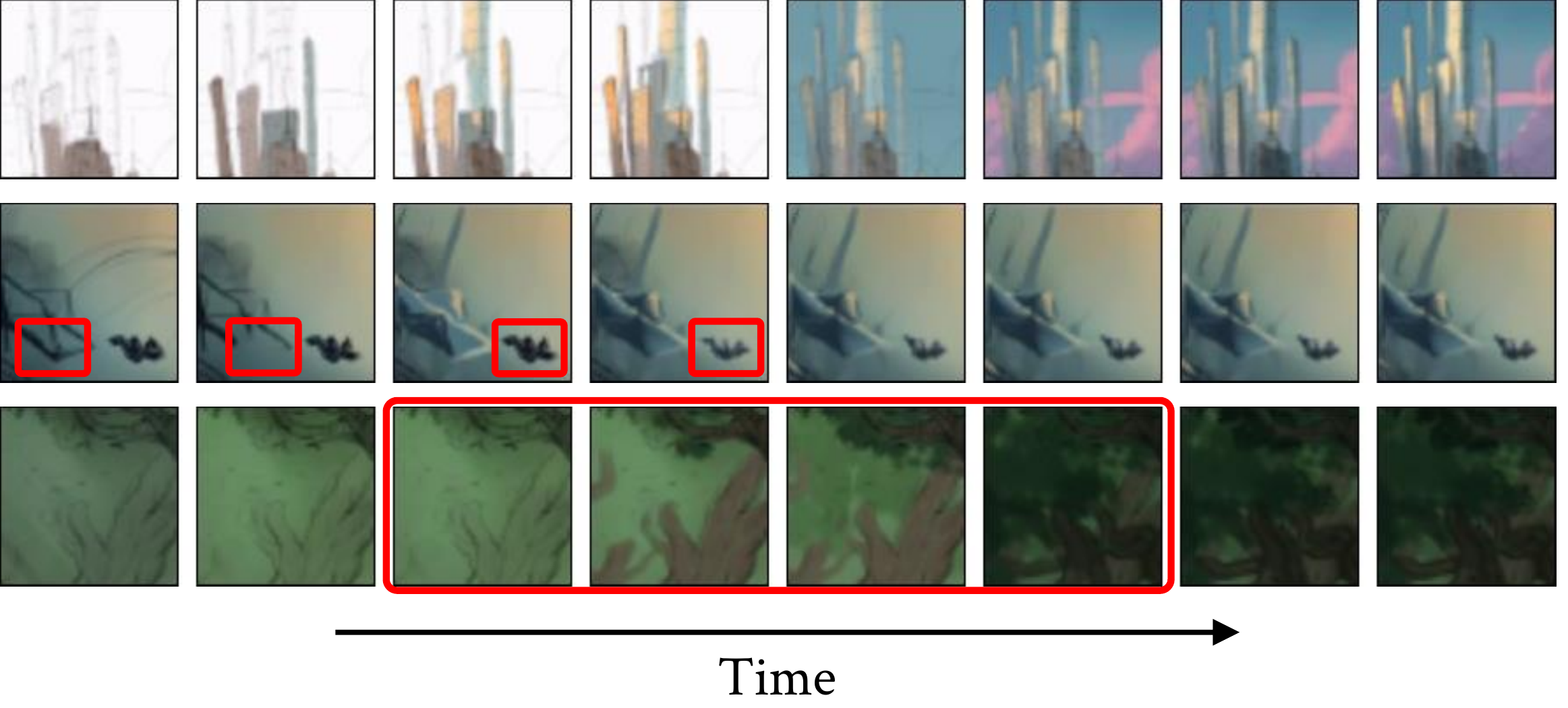}
\vspace{-18pt}
\caption{\textbf{Example digital painting sequences}. These sequences show a variety of ways to add paint, including fine strokes and filling (row 1), and broad strokes (row 3). We use red boxes to outline challenges, including erasing (row 2) and drastic changes in color and composition (row 3).}\label{fig:dataset_digital}
\end{figure}

\begin{figure}
\centering
\includegraphics[width=\linewidth]{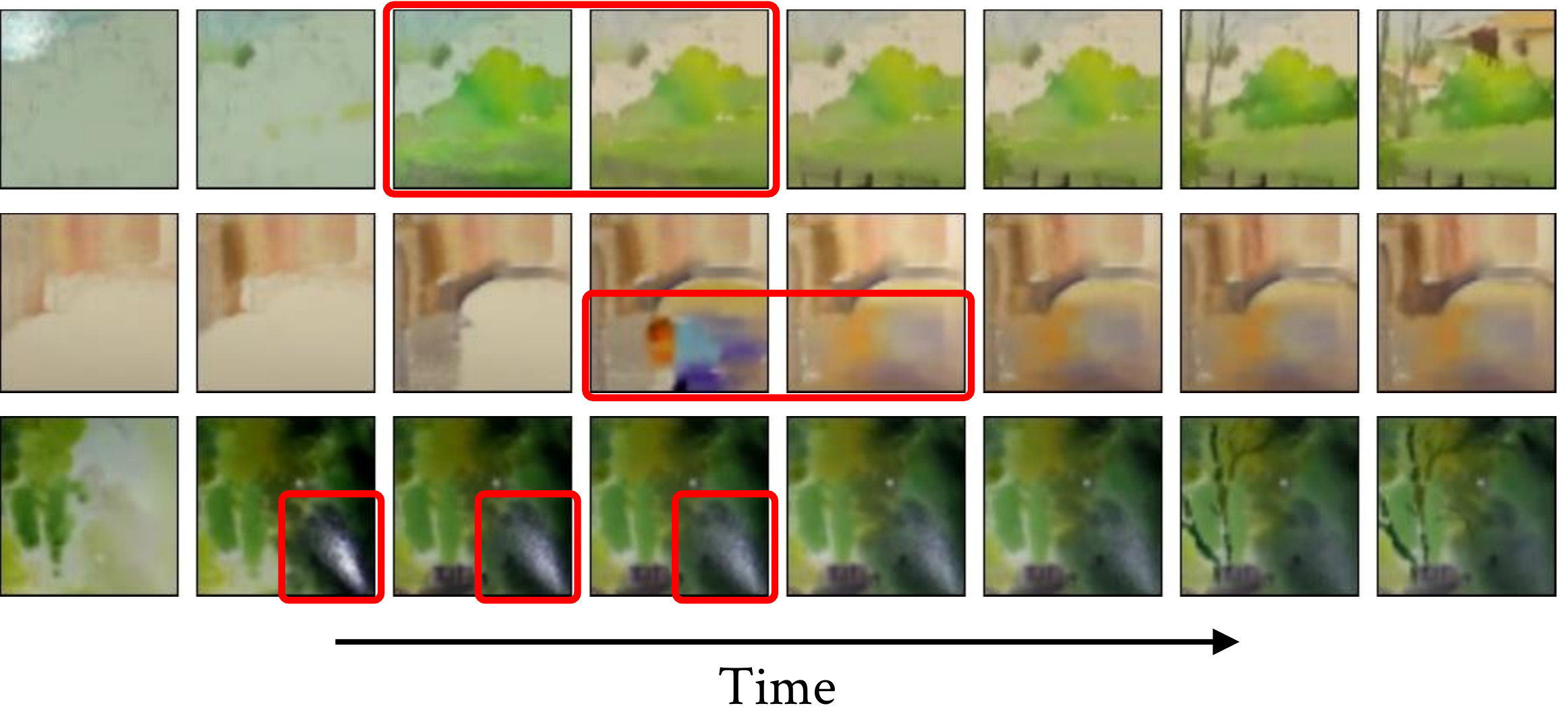}
\vspace{-18pt}
\caption{\textbf{Example watercolor painting sequences}. The outlined areas highlight some watercolor-specific challenges, including changes in lighting (row 1), diffusion and fading effects as paint dries (row 2), and specular effects on wet paint (row 3).}\label{fig:dataset_watercolors}
\end{figure}

\paragraph{High Variability}
\begin{enumerate}[topsep=2pt,itemsep=1pt,leftmargin=2ex]
\item[] \textbf{Painting trajectories}: Even for the same scene, different artists will likely paint objects in different temporal orders (Figure \ref{fig:problem_trajectories}).
\item[] \textbf{Painting rates}: Artists work at different speeds, and apply paint in different amounts.
\item[] \textbf{Scales and shapes}: Over the course of a painting, artists use strokes that vary in size and shape. Artists often use broad strokes early on, and add fine details later. 
\item[] \textbf{Data availability}: Due to the limited number of available videos in the wild, it is challenging to gather a dataset that captures the aforementioned types of variability.
\end{enumerate}
\vspace{-10pt}
\paragraph{Medium-specific challenges}
\begin{enumerate}[topsep=2pt,itemsep=1pt,leftmargin=2ex]
\item[] \textbf{Non-paint effects}: \hl{In digital art applications (\textit{e.g.}, \cite{procreate2019manual}), there are many} tools that apply local blurring, smudging, or specialized paint brush shapes. Artists can also apply global effects simulating varied lighting or tones. 
\item[] \textbf{Erasing effects}: In digital art applications, artists can erase or undo past actions, as shown in Figure \ref{fig:dataset_digital}.
\item[] \textbf{Physical effects in watercolor paintings}: Watercolor painting videos exhibit distinctive effects resulting from the physical interaction of paint, water, and paper. These effects include specular lighting on wet paint, pigments fading as they dry, and water spreading from the point of contact with the brush (Figure \ref{fig:dataset_watercolors}).
\end{enumerate}
In this work, we design a learning-based model to handle the challenges of high variability and painting medium-specific effects.

%% file: method.tex
\section{Method}\label{sec:timelapse_method}
We begin by formalizing the time lapse video synthesis problem. Given a painting $x_T$, our task is to synthesize the past frames $x_1,\cdots,x_{T-1}$. Suppose we have a training set of real time lapse videos $\{\mathbf{x}^{(i)}=x^{(i)}_1, \cdots, x^{(i)}_{T^{(i)}}\}$. We first define a principled probabilistic model, and then learn its parameters using these videos. At test time, given a completed painting, we sample from the model to create new videos \hl{of} realistic-looking painting processes.

\subsection{Model}\label{sec:timelapse_model}
We propose a probabilistic, temporally recurrent \hl{model for changes made during the painting process}. At each time instance $t$, the model predicts a pixel-wise intensity change $\delta_t$ that should be added to the previous frame to produce the current frame; that is, $x_t = x_{t-1} + \delta_t$. This change \hl{could represent one or multiple physical or digital paint strokes, or other effects such as erasing or fading.}

\begin{figure}[t]
    \centering
	\vspace{-8pt}
                \includegraphics[width=0.8\linewidth]{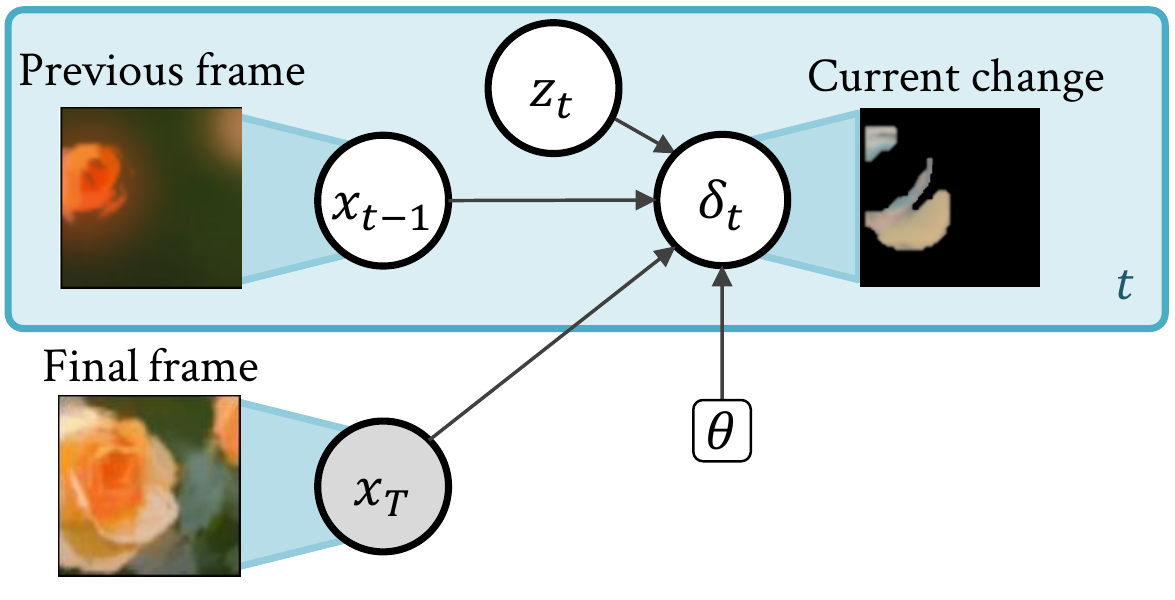}
                \vspace{-5pt}
        \caption{\textbf{The proposed probabilistic model}. Circles represent random variables; the shaded circle denotes a variable that is observed at inference time. The rounded rectangle represents model parameters.}
        \label{fig:plate_painter}
\end{figure}

We model $\delta_t$ as being generated from a random latent variable $z_t$, the completed piece $x_T$, and the image content at the previous time step $x_{t-1}$; the likelihood is $p_{\theta}(\delta_t|z_t,x_{t-1};x_T)$ (\hl{Figure \ref{fig:plate_painter}}). Using a random variable $z_t$ helps to capture the stochastic nature of painting. Using both $x_T$ and $x_{t-1}$ enables the model to capture time-varying effects such as the progression of coarse to fine brush sizes, while the Markovian assumption facilitates learning from a small number of video examples.

It is common to define such image likelihoods as a per-pixel normal distribution, which results in an L2 image similarity loss term in maximum likelihood formulations \cite{kingma2013auto}. In synthesis tasks, using L2 loss often produces blurry results~\cite{isola2016image}. We instead \hl{design our image similarity loss as the} L1 distance in pixel space and the L2 distance in a perceptual feature space. Perceptual losses are commonly used in image synthesis and style transfer tasks to produce sharper and more visually pleasing results~\cite{dosovitskiy2016generating,isola2016image,johnson2016perceptual,niklaus2017video,zhang2018unreasonable}. We use the L2 distance between normalized VGG16 features \cite{simonyan2014very} as described in \cite{zhang2018unreasonable}. We let the likelihood take the form:

\vspace{-15pt}
\small
\begin{align}
&p_{\theta}(\delta_t|z_t,x_{t-1};x_T)\nonumber\\
&\propto e^{-\frac{1}{\sigma_1}|\delta_t - \hat{\delta}_t|}\mathcal{N}\big(V(x_{t-1} + \delta_t); V(x_{t-1} + \hat{\delta}_t), \sigma_2^2\mathbbm{I}\big),\label{eq:prob_form_likelihood}
\end{align}
\normalsize
where $\hat{\delta}_t=g_{\theta}(z_t,x_{t-1},x_T)$, $g_{\theta}(\cdot)$ represents a function parameterized by $\theta$, $V(\cdot)$ is a function that extracts normalized VGG16 features, and $\sigma_1,\sigma_2$ are fixed noise parameters.

We assume the latent variable $z_t$ is generated from the multivariate standard normal distribution:  
\vspace{-6pt}
\begin{align}
p(z_t)&=\mathcal{N}(z_t; 0, \mathbbm{I}).\label{eq:prob_form_prior}
\end{align}

We aim to find model parameters $\theta$ that best explain all videos in our dataset:
\vspace{-2pt}
\begin{align}
&\argmax_\theta \Pi_i \Pi_t p_\theta(\delta^{(i)}_t,x^{(i)}_{t-1}, x^{(i)}_{T^{(i)}})\nonumber\\\vspace{-5pt}
&=\argmax_\theta \Pi_i \Pi_t \int_{z_t}p_\theta(\delta_t^{(i)}|z_t^{(i)},x_{t-1}^{(i)};x_{T^{(i)}}^{(i)})dz_t.\label{eq:optimization}
\end{align}
This integral is intractable, and the posterior $p(z_t|\delta_t,x_{t-1};x_T)$ is also intractable, preventing the use of the EM algorithm. We instead use variational inference and introduce an approximate posterior distribution $p(z_t|\delta_t,x_{t-1};x_T)\approx q_\phi(z_t|\delta_t,x_{t-1};x_T)$ \cite{kingma2014semi,xue2016visual,yan2016attribute2image}. We let this approximate distribution take the form of a multivariate normal:
\begin{align}
&q_\phi(z_t|\delta_t,x_{t-1}, x_T)\nonumber\\
&\hspace{5pt}=\mathcal{N}\big(z_t; \mu_\phi(\delta_t,x_{t-1},x_T), \Sigma_\phi(\delta_t,x_{t-1},x_T)\big),\label{eq:prob_form_posterior}
\end{align}
where $\mu_\phi(\cdot),\Sigma_\phi(\cdot)$ are functions parameterized by $\phi$, and $\Sigma_\phi(\cdot)$ is diagonal.

\subsubsection{Neural network framework}
\begin{figure}[t]
\centering
\vspace{-8pt}
\includegraphics[width=\linewidth]{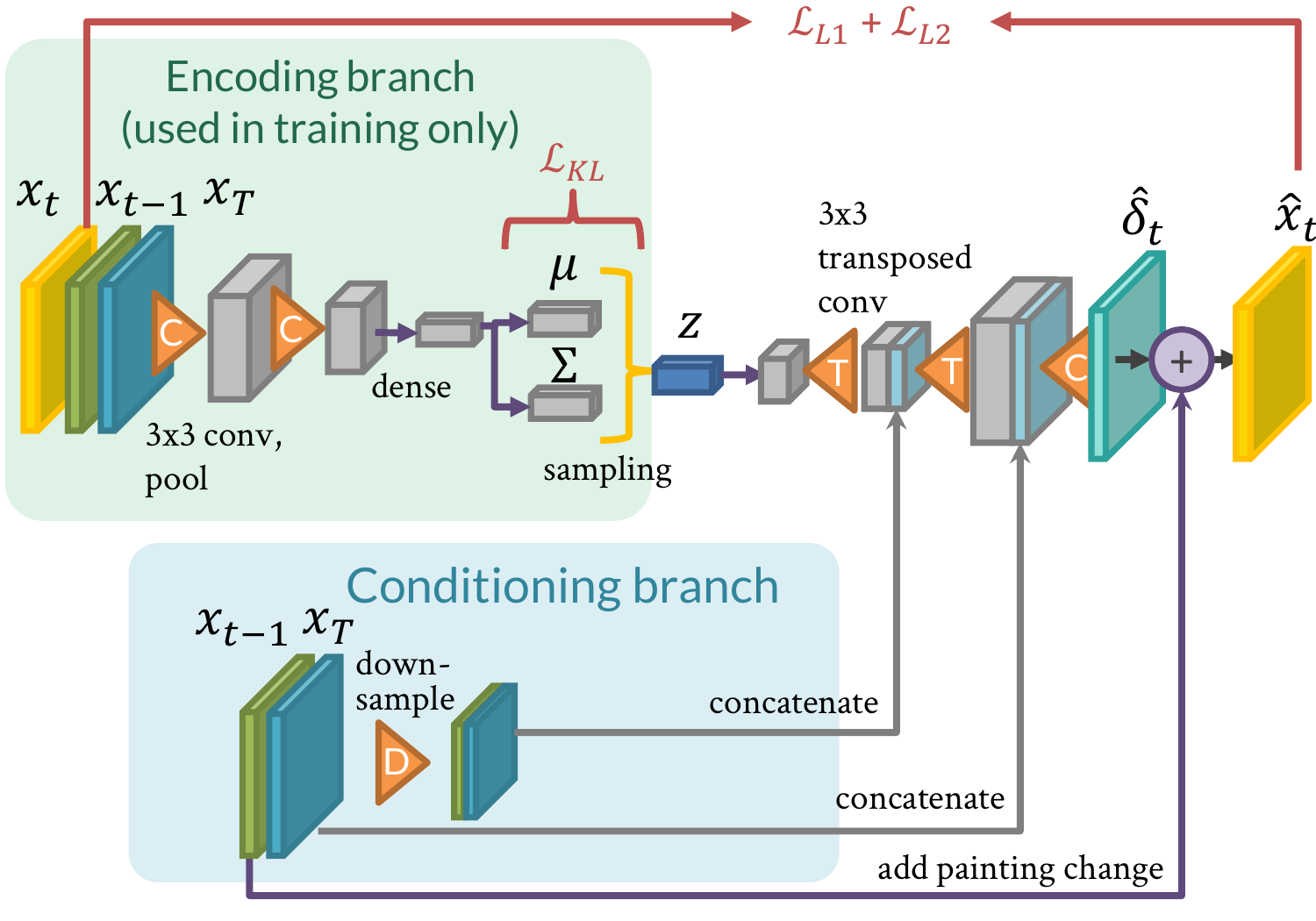}\vspace{-6pt}
\caption{\textbf{Neural network architecture}. We implement our model using a conditional variational autoencoder framework. At training time, the network is encouraged to reconstruct the current frame $x_t$, while sampling the latent $z_t$ from a distribution that is close to the standard normal. At test time, the encoding branch is removed, and $z_t$ is sampled from the standard normal. We use the shorthand ${\hat{\delta}_t=g_\theta(z_t, x_{t-1},x_T)},{\hat{x}_t=x_{t-1} + \hat{\delta}_t}$.}\label{fig:network_cvae}
\end{figure}
We implement the functions $g_\theta$, $\mu_\phi$ and $\Sigma_\phi$ as a convolutional encoder-decoders parameterized by $\theta$ and $\phi$, using a conditional variational autoencoder (CVAE) framework~\cite{walker2016uncertain,yan2016attribute2image}. We use an architecture similar to \cite{yan2016attribute2image}. We summarize our architecture in Figure \ref{fig:network_cvae} and include full details in the appendix.

\subsection{Learning}\label{sec:learning}
We learn model parameters using short sequences from the training video dataset, which we discuss in further detail in Section \ref{sec:datasets}. We use two stages of optimization to facilitate convergence: \textit{pairwise optimization}, followed by \textit{sequence optimization}. 

\subsubsection{Pairwise optimization}\label{sec:optimization}
From Equations \eqref{eq:optimization} and \eqref{eq:prob_form_posterior}, we obtain an expression for each \textit{pair} of consecutive frames (a derivation is provided in the appendix):
\vspace{-5pt}
\begin{align}
&\log  p_\theta(\delta_t,x_{t-1}, x_T)\nonumber\\
&\geq\mathbb{E}_{z_t\sim q_\phi(z_t|x_{t-1}, \delta_t;x_T)}\big[\log p_\theta(\delta_t|z_t,x_{t-1};x_T)\big]\nonumber\\
&\hspace{10pt}- KL[q_\phi(z_t|\delta_t,x_{t-1};x_T)||p(z_t)],\label{eq:elbo}
\end{align}
where $KL[\cdot||\cdot]$ denotes the Kullback-Liebler divergence. 
Combining Equations \eqref{eq:prob_form_likelihood}, \eqref{eq:prob_form_prior}, \eqref{eq:prob_form_posterior}, and \eqref{eq:elbo}, we minimize:
\vspace{-15pt}
\begin{align}
&\mathcal{L}_{KL} + \frac{1}{\sigma_1} \mathcal{L}_{L1}(\delta_t, \hat{\delta}_t) \nonumber\\
&+\frac{1}{2\sigma_2^2} \mathcal{L}_{L2}(V(x_{t-1} + \delta_t), V(x_{t-1} + \hat{\delta}_t)),\label{eq:loss_pairwise}
\end{align}
where $\mathcal{L}_{KL}= \frac{1}{2}\big(- \log\Sigma_\phi + \Sigma_\phi + \mu_\phi^2\big)$, and \hl{the image similarity terms} $\mathcal{L}_{L1},\mathcal{L}_{L2}$ represent L1 and L2 distance respectively. 

We optimize Equation \eqref{eq:loss_pairwise} on single time steps, which we obtain by sampling all pairs of consecutive frames from the dataset. We also train the model to produce the first frame $x_1$ from videos that begin with a blank canvas, given a white input frame $x_{blank}$, and $x_T$. These \textit{starter sequences} teach the model how to start a painting at inference time. 

\begin{figure}
\vspace{-10pt}
\hspace{-10pt}
\includegraphics[width=1.04\linewidth]{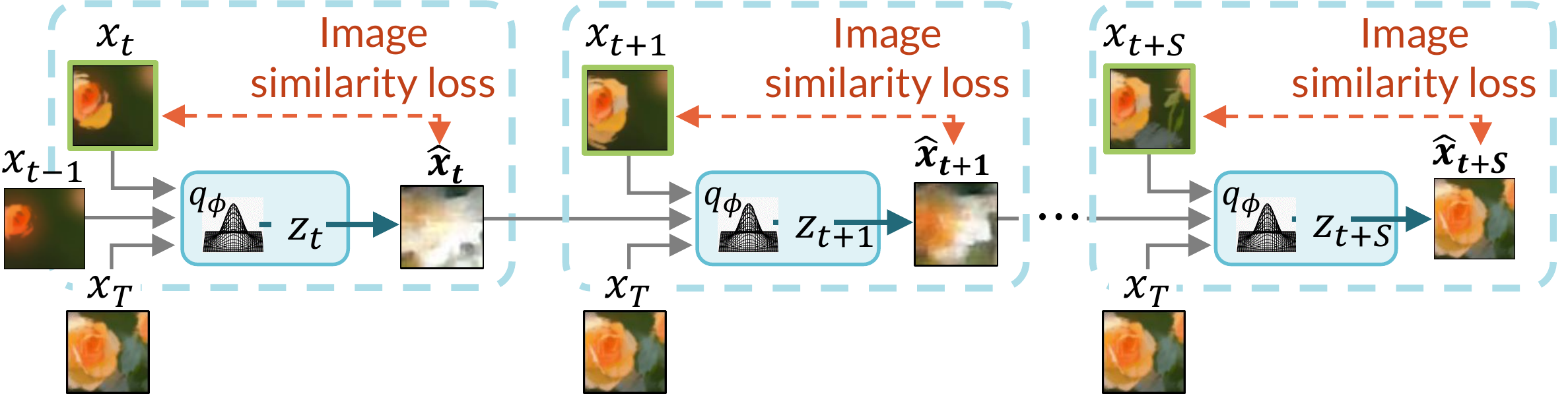}\vspace{-5pt}\caption{\textbf{Sequential CVAE training}. Our model is trained to reconstruct a real frame (outlined in green) while building upon its previous predictions for $S$ time steps.}\label{fig:training_seq_cvae}
\end{figure}

\subsubsection{Sequence optimization}
To synthesize an entire video, we run our model recurrently for multiple time steps, building upon its own predicted frames. It is common when making sequential predictions to observe compounding errors or artifacts over time~\cite{villegas2017learning}. We use a novel training scheme to encourage outputs of the model to be accurate and realistic over multiple time steps. We alternate between two sequential training modes.

\begin{figure}
\vspace{-10pt}
\hspace{-8pt}
\includegraphics[width=1.04\linewidth]{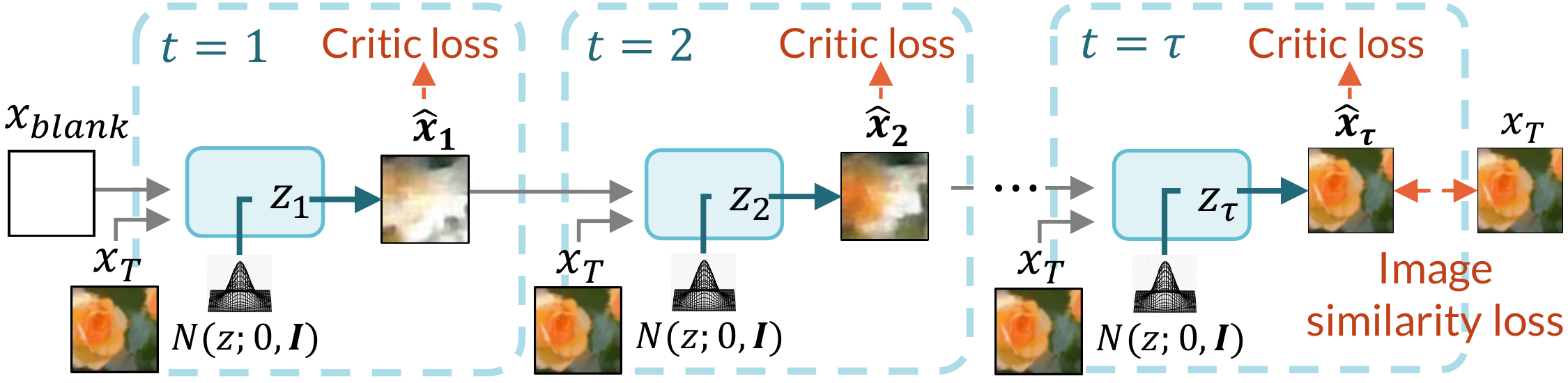}\caption{\textbf{Sequential sampling training}. We use a conditional frame critic to encourage all frames sampled from our model to look realistic. The image similarity loss on the final frame encourages the model to complete the painting in $\tau$ time steps.}\label{fig:training_seq}
\end{figure}

\begin{itemize}[leftmargin=0pt, itemsep=1pt]
\item[] \textbf{Sequential CVAE training} encourages \textit{sequences} of frames to be well-captured by the learned distribution, by reducing the compounding of errors. We train the model sequentially for several time steps, predicting each intermediate frame $\hat{x}_t$ using the model's \hl{prediction} \hl{from} the previous time step: \hl{$\hat{x}_t=\hat{x}_{t-1} + g_{\theta}(z_t,\hat{x}_{t-1},x_T)$ for \hbox{$z_t \sim q_\phi(z_t|x_t-\hat{x}_{t-1},\hat{x}_{t-1}, x_T)$}}. We compare each predicted frame to its corresponding real frame using the image similarity losses in Eq. \eqref{eq:loss_pairwise}. We illustrate this in Figure \ref{fig:training_seq_cvae}.

\item[] \textbf{Sequential sampling training} encourages random samples from our learned distribution to look like \textit{realistic} partially-completed paintings. During inference (described below), we rely on sampling from the prior $p(z_t)$ at each time step to synthesize new videos. A limitation of the variational strategy is the limited coverage of the latent space $z_t$ during training \cite{engel2018latent}, sometimes leading to predictions during inference $\hat{x}_t=\hat{x}_{t-1} + g_{\theta}(z_t,\hat{x}_{t-1},x_T)$, $z_t\sim p(z_t)$ that are unrealistic. To compensate for this, we introduce supervision on such samples by amending the \hl{image similarity} term in Equation \eqref{eq:elbo} with a conditional critic loss term~\cite{gulrajani2017improved}:
\begin{align}
\mathcal{L}_{critic}=&\mathbb{E}_{z_t\sim p(z_t)}\big[D_\psi\big(\hat{x}_t,\hat{x}_{t-1},x_T\big)\big]\nonumber\\
&- \mathbb{E}_{x_t}\big[D_\psi(x_t,x_{t-1},x_T)\big],
\end{align}
where $D_\psi(\cdot)$ is a critic function with parameters $\psi$. This critic encourages the distribution of sampled \hl{changes} ${\hat{\delta}_t=g_\theta(z_t,\hat{x}_{t-1},x_T),z_t \sim p(z_t)}$ to match the distribution of training \hl{painting changes} $\delta_t$. We use a critic architecture based on~\cite{choi2018stargan} and optimize it using WGAN-GP~\cite{gulrajani2017improved}.

In addition to the critic loss, we apply the image similarity losses (discussed above) after $\tau$ time steps, to encourage the model to eventually produce the completed painting. \hl{This training scheme is summarized in Figure \ref{fig:training_seq}.}
\end{itemize}

\subsection{Inference: video synthesis}\label{sec:inference}
Given a completed painting $x_T$ and learned model parameters $\theta,\phi$, we synthesize videos by sampling from the model at each time step. Specifically, we synthesize each frame $\hat{x}_t=\hat{x}_{t-1}+g_\theta(z_t,\hat{x}_{t-1},x_T)$ using the synthesized previous frame $\hat{x}_{t-1}$ and a randomly sampled $z_t \sim p(z_t)$. We start each video using $\hat{x}_{0}=x_{blank}$, a blank frame.

\subsection{Implementation}
We implement our model using Keras \cite{chollet2015} and Tensorflow \cite{abadi2016}. We experimentally selected the hyperparameters controlling the reconstruction loss weights to be $\sigma_1=0.01$ and $\sigma_2=0.1$, using the validation set.

%% file: experiments.tex
\section{Experiments}\label{sec:timelapse_experiments}
\subsection{Datasets}\label{sec:datasets}
We collected \hl{time lapse recordings} of paintings from YouTube and Vimeo. We selected digital and watercolor paintings (which are common painting methods on these websites), and focused on landscapes or still lifes (which are common subjects for both mediums). We downloaded each video at $360\times 640$ resolution and cropped it temporally and spatially to include only the painting process (excluding \hl{other content such as introductions or sketching}). We split each dataset in a 70:15:15 ratio into training, validation, and held-out test video sets.

\begin{enumerate}[itemsep=1pt,leftmargin=0pt]
\item[] \textbf{Digital paintings:}
We collected $117$ digital painting time lapses. The average duration is 4 minutes, with many videos having already been sped up by artists using the Procreate application \cite{procreate2019manual}. We selected videos with minimal zooming and panning. We manually removed segments that contained movements such as translations, flipping and zooming. Figure \ref{fig:dataset_digital} shows example video sequences. 

\item[] \textbf{Watercolor paintings:}
We collected $116$ watercolor time lapses, with an average duration of 20 minutes. We only kept videos that contained minimal movement of the paper, and manually corrected any small translations of the painting. We show examples in Figure \ref{fig:dataset_watercolors}. 

A challenge with videos of physical paintings is the presence of the hand, paintbrush and shadows in many frames. We trained a simple convolutional neural network to identify and remove frames that contained these artifacts. 
\end{enumerate}
\subsubsection{Sequence extraction}\label{sec:seq_criteria}
We synthesize time lapses at a lower temporal resolution than real-time for computational feasibility. We extract training sequences from raw videos at a period of $\gamma > 0$ frames (\textit{i.e.,} skipping $\gamma$ frames in each synthesized time step), with a maximum variance of $\epsilon$ frames. \hl{Allowing some variance in the sampling rate is useful for (1) improving robustness to varied painting rates, and (2) extracting sequences from watercolor painting videos where many frames containing hands or paintbrushes have been removed}. We select $\gamma$ and $\epsilon$ independently for each dataset. We avoid capturing static segments of each video (\textit{e.g.,} when the artist is speaking) by requiring that adjacent frames in each sequence have at least $1\%$ of the pixels changing by a fixed intensity threshold. We use a dynamic programming method to find all \hl{training and validation} sequences that satisfy these criteria. We train on sequences of length 3 or 5 for sequential CVAE training, and length $\tau=40$ for sequential sampling training, which we determined using experiments on the validation set. For \hl{evaluations on the} test set, we extract a single sequence from each test video that satisfies the filtering criteria.

\subsubsection{Crop extraction}
To facilitate learning from small numbers of videos, we use multiple crops from each video. We first downsample each video spatially to $126 \times 168$, so that most patches contain visually interesting content and spatial context, \hl{and then extract $50\times 50$ crops with minimal overlap}. 

\subsection{Baselines}
\begin{itemize}[itemsep=1pt,topsep=2pt,leftmargin=0pt]
\item[] \textbf{Deterministic video synthesis (\textit{unet}):}
In image synthesis tasks, it is common to use an encoder-decoder architecture with skip connections, similar to U-Net \cite{isola2016image,ronneberger2015u}. We adapt this technique to synthesize an entire video at once.  

\item[] \textbf{Stochastic video synthesis (\textit{vdp}):}
Visual deprojection synthesizes a distribution of videos from a single temporally-projected input image \cite{balakrishnan2019visual}.
\end{itemize}
\vspace{5pt}
We design each baseline model architecture to have a comparable number of parameters to our model. Both baselines output videos of a fixed length, which we choose to be $40$ to be comparable to our choice of $\tau=40$ in Section \ref{sec:seq_criteria}.

\subsection{Results}\label{sec:results}
We conducted both quantitative and qualitative evaluations. We first present a user study quantifying human perception of the realism of our synthesized videos. Next, we qualitatively examine our synthetic videos, and discuss characteristics that contribute to their realism. Finally, we discuss quantitative metrics for comparing sets of sampled videos to real videos. \hl{We show additional results, including videos and visualizations using the tipiX tool \cite{dalcatipix} on our project page at \texttt{https://xamyzhao.github.io/timecraft}.}

We experimented with training each method on digital or watercolor paintings only, as well as on the combined paintings dataset. For all methods, we found that training on the combined dataset produced the best qualitative and quantitative results (likely due to our limited dataset size), and we only present results for those models.

\subsubsection{Human evaluations}
We surveyed 158 people using Amazon Mechanical Turk~\cite{mturk}. Participants compared the realism \hl{of pairs of videos randomly sampled} from \textit{ours}, \textit{vdp}, or the real videos. In this study, we omit the weaker baseline \textit{unet}, which performed consistently worse on all metrics (discussed below). 

We first trained the participants by showing them several examples of real painting time lapses. We then presented a pair of time lapse videos generated by different methods for the center crop of the same painting, and asked ``Which video in each pair shows a more realistic painting process?'' We repeated this process for 14 randomly sampled paintings from the test set. Full study details are in the appendix. 

Table \ref{tab:results_mturk} indicates that almost every participant thought videos \hl{synthesized} by our model looked more realistic than those \hl{synthesized} by \textit{vdp} ($p < 0.0001$). Furthermore, participants confused our synthetic videos with real videos nearly half of the time. In the next sections, we show example synthetic videos and discuss aspects that make \hl{our model's results} appear more realistic, offering an explanation for these promising user study results.
\begin{table}[t]
\centering
\hspace{-3pt}
\begin{tabular}{M{1.6cm}|M{1.7cm}|M{1.7cm}|M{1.7cm}}
Comparison&All paintings & Watercolor paintings & Digital paintings\\
\hline
real $>$ \textit{vdp}&90\% & 90\% & 90\%\\
real $>$ \textit{ours}&55\%& 60\% & 51\%\\
\textit{ours} $>$ \textit{vdp}&91\%& 90\% & 88\%\\
\end{tabular}\caption{\small \textbf{User study results.} Users compared the realism of pairs of videos randomly sampled from \textit{ours}, \textit{vdp}, and real videos. The vast majority of participants preferred our videos over \textit{vdp} videos ($p < 0.0001$). Similarly, most participants chose real videos over \textit{vdp} videos ($p < 0.0001$). Users preferred real videos over ours ($p = 0.0004$), but many participants confused our videos with the real videos, especially for digital paintings.}\label{tab:results_mturk}
\end{table}
\normalsize
\subsubsection{Qualitative results}\label{sec:results_qual}
\begin{figure}[t]
\centering
\subfloat{
\includegraphics[width=0.95\linewidth]{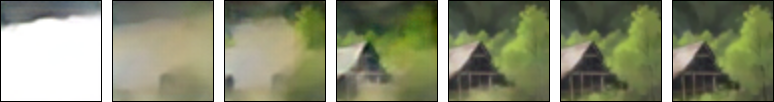}}\\\vspace{-8pt}
\subfloat{
\includegraphics[width=0.95\linewidth]{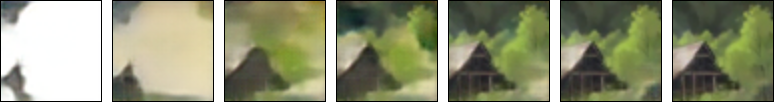}}\\\vspace{-8pt}
\subfloat{
\includegraphics[width=0.95\linewidth]{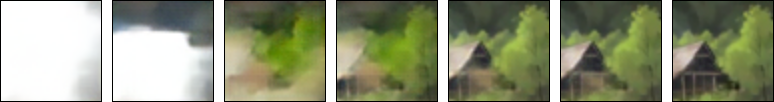}}\\\vspace{-5pt}
\subfloat{
\includegraphics[width=0.95\linewidth]{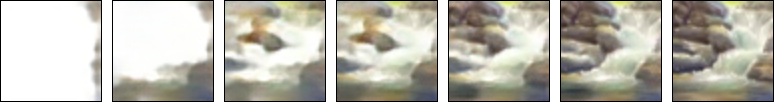}}\\\vspace{-8pt}
\subfloat{
\includegraphics[width=0.95\linewidth]{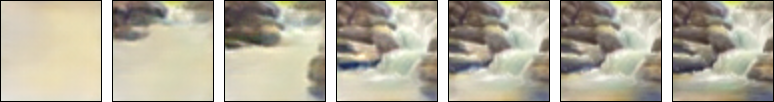}}\vspace{-8pt}
\subfloat{
\includegraphics[width=0.95\linewidth]{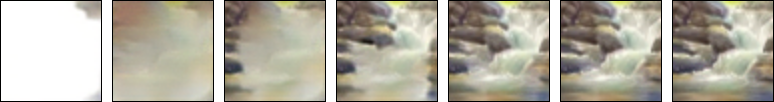}}\\\vspace{2pt}
\caption{\textbf{Diversity of sampled videos}. We show examples of our method applied to a digital (top 3 rows) and a watercolor (bottom 3 rows) painting from the test set. Our method captures diverse and plausible painting trajectories. }
\label{fig:results_samples}
\end{figure}

Figure \ref{fig:results_samples} shows sample sequences produced by our model for two input paintings. Our model chooses different orderings of semantic regions \hl{during the painting process}, leading to different paths that still converge to the same completed painting. 

\begin{figure*}[h]
\centering
\vspace{-18pt}
\subfloat[\textbf{Similarly to the artist, our method paints in a coarse-to-fine manner}. Blue arrows show where our method first applies a flat color, and then adds fine details. Red arrows indicate where the baselines add fine details even in the first time step.]{{
\includegraphics[width=0.9\linewidth]{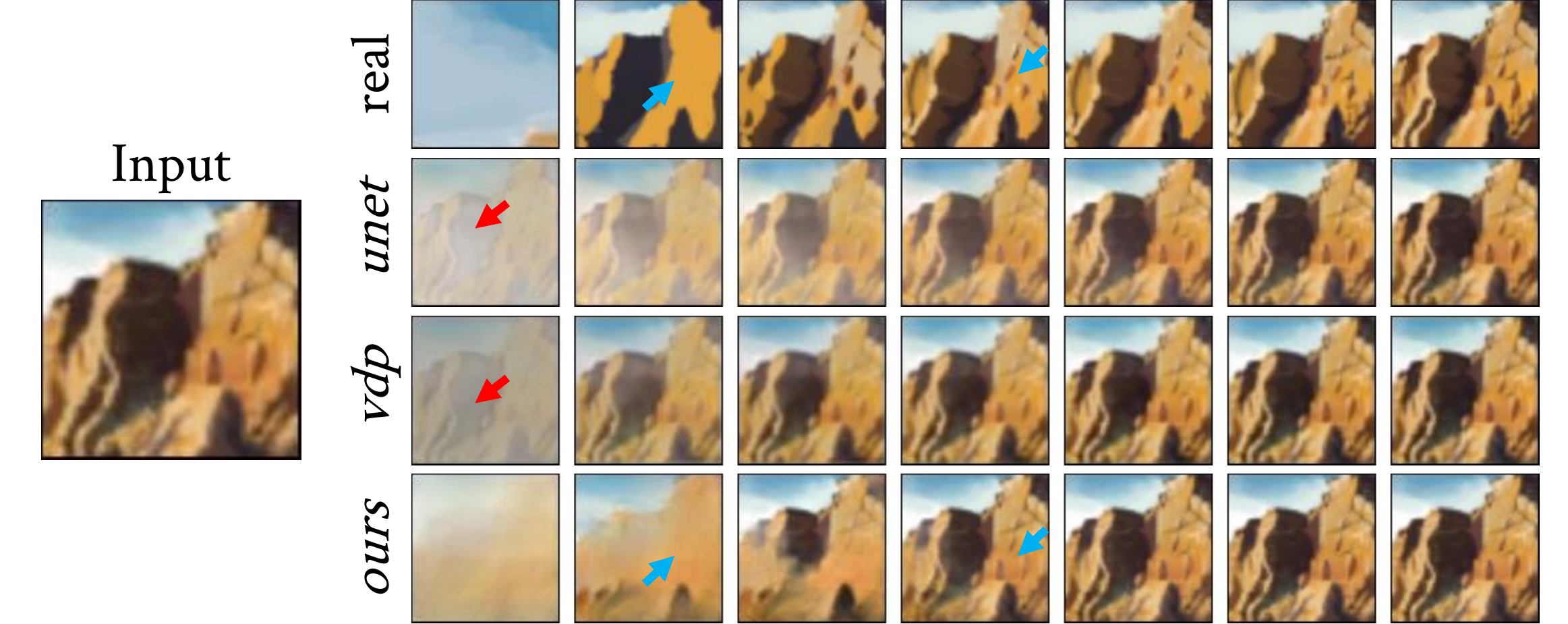}
}}\\
\subfloat[\textbf{Our method works on similar regions to the artist, although it does not use the same color layers to achieve the completed painting}. Blue arrows show where our method paints similar parts of the scene to the artist (filling in the background first, and then the house, and then adding details to the background). Red arrows indicate where the baselines do not paint according to semantic boundaries, gradually fading in the background and the house in the same time step.]{{
\includegraphics[width=0.9\linewidth]{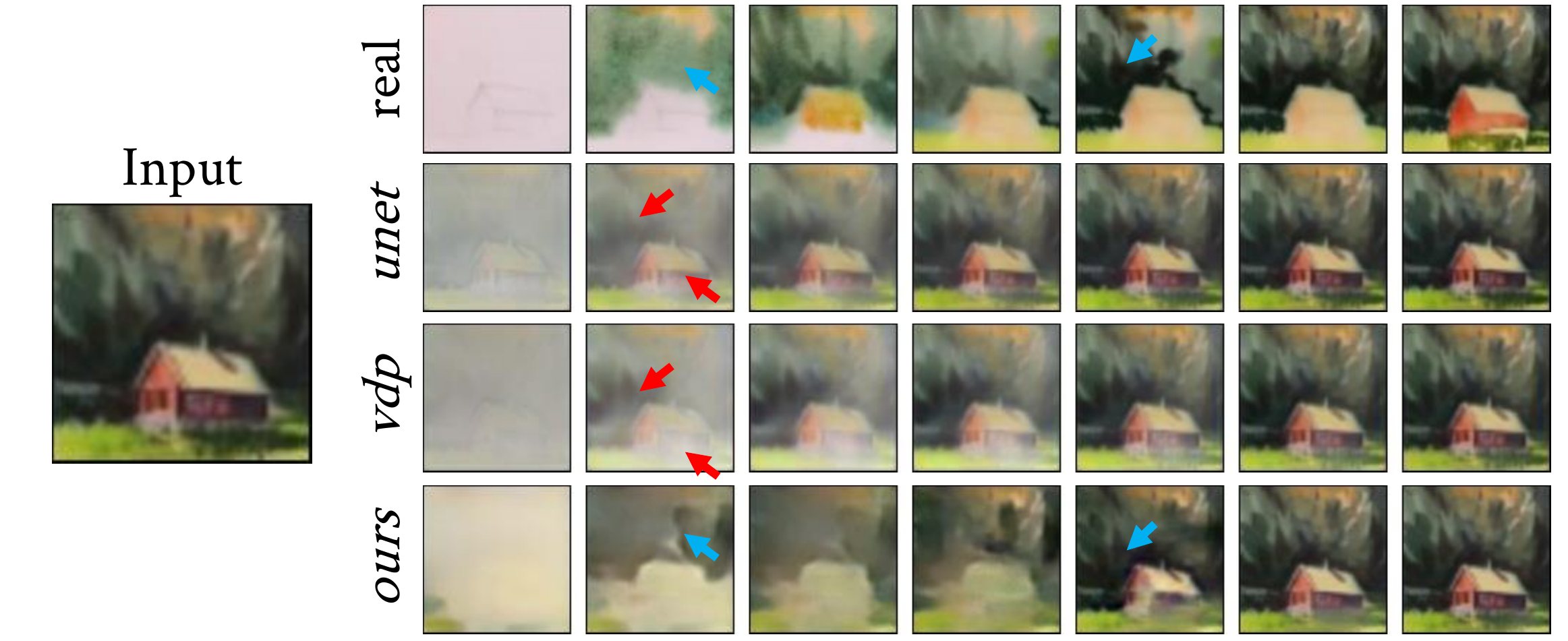}
}}\caption{\textbf{Videos predicted from the digital (top) and watercolor (bottom) test sets}. For the stochastic methods \textit{vdp} and \textit{ours}, we show the nearest sample to the real video out of 2000 samples. We show additional results in the appendix.}\label{fig:results_detailed}
\normalsize
\end{figure*}

Figure \ref{fig:results_detailed} shows \hl{videos} synthesized by each method. To objectively compare the stochastic methods \textit{vdp} and \textit{ours}, we show the most similar \hl{predicted video} by L1 distance to the ground truth video. The ground truth \hl{videos} show that artists tend to paint in a coarse-to-fine manner, using broad strokes near the start of a painting, and finer strokes near the end. \hl{Artists also tend to focus on one or a few semantic regions in each time step}. As we highlight with arrows, our method captures \hl{these tendencies} better than baselines, having learned to \hl{make changes within} separate semantic regions such as mountains, cabins and trees. Our predicted trajectories are similar to the ground truth, showing that our sequential modeling approach is effective at capturing realistic temporal progressions. \hl{In contrast, the baselines tend to make blurry changes without separating the scene into components.}

\begin{figure*}[t]
\arxiv{
\centering
\subfloat[\textbf{The proposed method does not always synthesize realistic \hl{changes} for fine details}. Blue arrows highlight frames where the method makes realistic \hl{painting changes}, working in one or two semantic regions at a time. Red arrows show examples where our method sometimes fills in many details in the frame at once.]{
\includegraphics[width=0.9\linewidth]{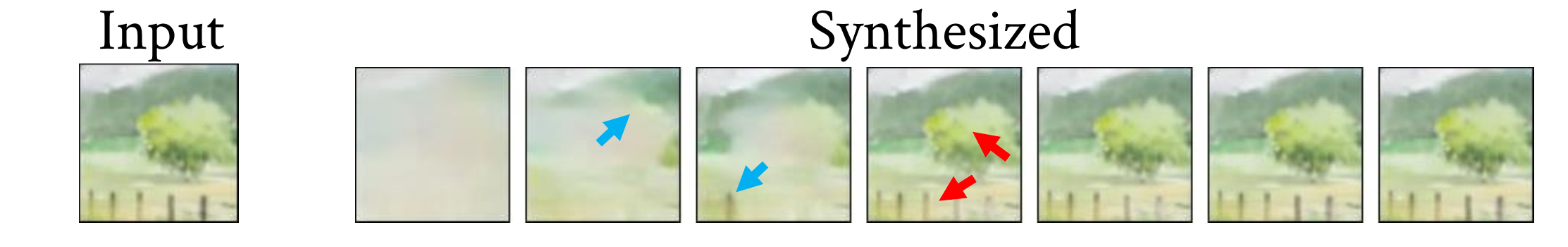}}\vspace{-5pt}\\
\subfloat[\textbf{The proposed method sometimes synthesizes changes in disjoint regions}. Red arrows indicate where the method produces \hl{painting changes} that fill in small patches that correspond to disparate semantic regions, leaving unrealistic blank gaps throughout the frame. This example also fills in much of the frame in one time step, although most of the filled areas in the second frame are coarse.]{
\includegraphics[width=0.9\linewidth]{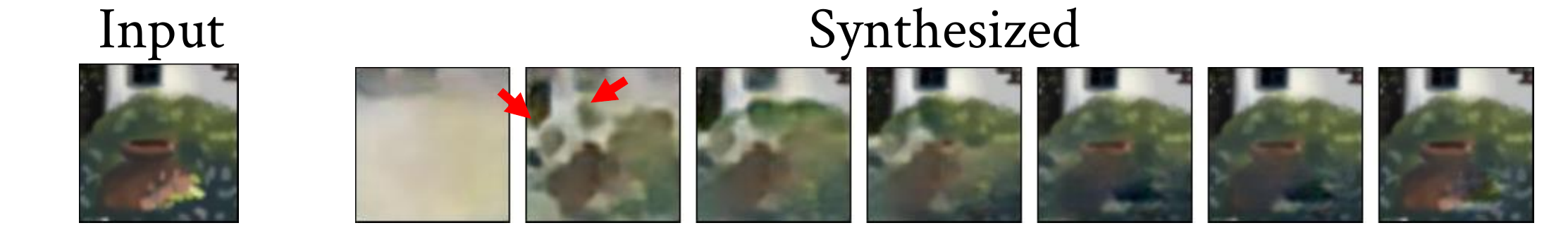}
}\caption{\arxiv{\textbf{Failure cases}. We show unrealistic effects that are sometimes synthesized by our method, for a watercolor painting (top) and a digital painting (bottom).}}\label{fig:failures}}
\end{figure*}

\arxiv{
We examine failure cases from the proposed method in Figure \ref{fig:failures}, such as making many fine or disjoint changes in a single time step and creating an unrealistic effect.}

\begin{figure*}[t]
\arxiv{
\centering
\subfloat[Digital paintings test set.]{
\includegraphics[trim=0 0 210 0,clip,scale=0.45]{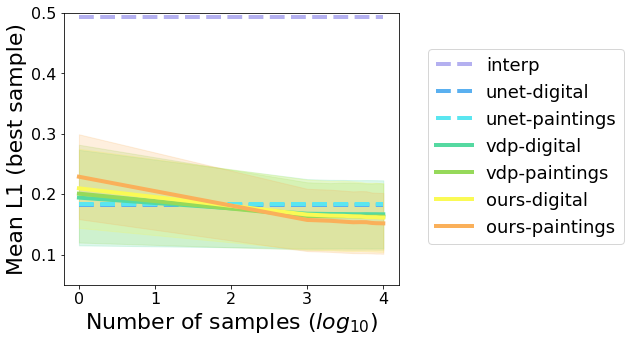}~
\includegraphics[scale=0.45]{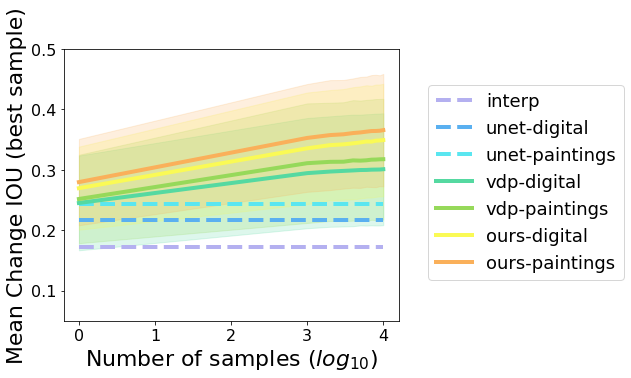}
}\\
\subfloat[Watercolor paintings test set.]{
\includegraphics[trim=0 0 228 0,clip,scale=0.45]{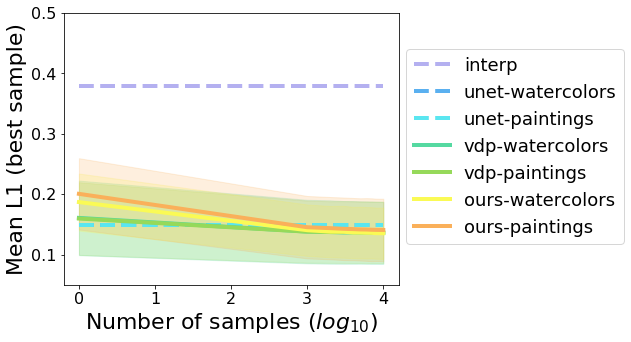}~
\includegraphics[scale=0.45]{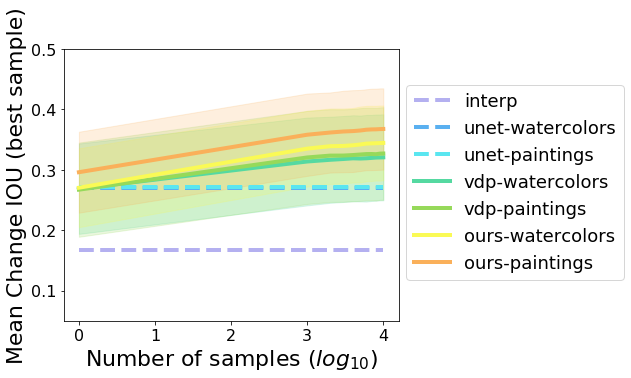}
}
\caption{\arxiv{\textbf{Quantitative results for varying numbers of samples}. As we draw more samples from each stochastic method (solid lines), the best video similarity to the real video improves. This indicates that some samples are close to the artist's specific painting choices. We use L1 distance as the metric on the left (lower is better), and \hl{change} IOU on the right (higher is better). Shaded regions show standard deviations of the stochastic methods. We highlight several insights from these plots. (1) Both our method and \textit{vdp} produce samples that are comparably similar to the real video by L1 distance (left). However, our method synthesizes \hl{painting changes} that are more similar in shape to those used by artists (right). (2) At low numbers of samples, the deterministic \textit{unet} method is closer (by L1 distance) to the real video than samples from \textit{vdp} or \textit{ours}, since L1 favors blurry frames that average many possibilities. (3) Our method shows more improvement in L1 distance and \hl{painting change} IOU than \textit{vdp} as we draw more samples, indicating that our method captures a more varied distribution of videos.}}\label{fig:plots}}
\end{figure*}


\subsubsection{Quantitative results}
\hl{In a stochastic task}, comparing synthesized results to ``ground truth'' is ill-defined, and developing quantitative measures of realism is difficult \cite{isola2017image,salimans2016improved}; these challenges motivated our user study above. In this section, we explore quantitative metrics designed to measure aspects of time lapse realism. For each video in the test set, we extract a 40-frame long sequence according to the criteria described in Section \ref{sec:seq_criteria}, and evaluate each method on 5 random crops using several video similarity metrics:

\begin{itemize}[itemsep=1pt,leftmargin=0pt]
\item[] \textbf{Best (across $k$ samples) overall video \hl{distance (lower is better)}:} For each \hl{crop}, we draw $k$ sample videos from each model and report the closest sample to the true video \hl{by L1 distance}~\cite{balakrishnan2019visual}. \arxiv{If a method has captured the distribution of real time lapses well, it should produce better ``best'' estimates as $k \rightarrow \infty$.} This captures whether \hl{a model produces some realistic samples, and whether the model is diverse enough to capture each artist's specific choices}. 

\item[] \textbf{Best (across $k$ samples) \hl{painting change} shape similarity \hl{(higher is better)}:} We quantify how similar the set of \hl{painting change} shapes are between the ground truth and each predicted video, disregarding the order in which they were performed. We define the \textit{\hl{painting change} shape} as a binary map of the changes made in each time step. For \hl{each time step in each test video}, we compare the artist's \hl{change} shape to the most similarly shaped \hl{change} synthesized by each method, as measured by intersection-over-union (IOU). This captures whether a method paints in similar semantic regions to the artist. 
\end{itemize}

\hl{We summarize these results in Table \ref{tab:results_quantitative}}. We include a deterministic \textit{interp} baseline, which linearly interpolates in time, as a quantitative lower bound. The deterministic \textit{interp} and \textit{unet} approaches perform poorly for both metrics. \hl{For $k=2000$,} \textit{vdp} and our method produce samples that lead to comparable ``best video similarity'' by L1 distance, highlighting the strength of methods designed to capture distributions of videos. The \hl{painting change} IOU metric shows that our method synthesizes \hl{changes} that are significantly more realistic than the other methods. 

\arxiv{We show the effect of increasing the number of samples $k$ in Figure \ref{fig:plots}. At low $k$, the blurry videos produced by \textit{interp} and \textit{unet} attain lower L1 distances to the real video than the videos produced by \textit{vdp} and \textit{ours} do, likely because L1 distance penalizes samples with different painting progressions more than it penalizes blurry ``average'' frames. In other words, an artist's time lapse will typically have a higher L1 distance to a video of a different but plausible painting process, than it would to a blurry, gradually fading video with ``average'' frames. As $k$ increases, \textit{vdp} and our method produce some samples that are close to the real video. Together with the user study described above, these metrics indicate that our method can capture a realistic variety of painting time lapses.}
\begin{table}[t]
\smaller
\centering
\hspace{-14pt}
\begin{tabular}{M{0.73cm}|M{1.43cm}|M{1.52cm}|M{1.43cm}|M{1.52cm}}
\multirow{2}{*}{Method}&\multicolumn{2}{c}{Digital paintings}&\multicolumn{2}{c}{Watercolor paintings}\\
 &L1&\hl{Change} IOU&L1&\hl{Change} IOU\\
\hline
interp&$0.49$ $(0.13)$&${0.17}$ $(0.06)$&$0.38$ $(0.09)$&${0.17}$ $(0.09)$\\
unet&$0.18$ $(0.08)$&$0.24$ $(0.08)$&$0.15$ $(0.06)$&$0.27$ $(0.07)$\\
vdp&$\mathbf{0.16}$ $(0.06)$&$0.31$ $(0.10)$&$\mathbf{0.14}$ $(0.05)$&$0.32$ $(0.08)$\\
ours&$\mathbf{0.16}$ $(0.05)$&$\mathbf{0.36}$ $(0.09)$&$\mathbf{0.14}$ $(0.05)$&$\mathbf{0.36}$ $(0.07)$\\
\end{tabular}\caption{\textbf{Quantitative results}. We compare videos synthesized from the digital and watercolor painting test sets to the artists' videos. For the stochastic methods \textit{vdp} and \textit{ours}, we draw 2000 video samples and report the closest one to the ground truth.}\label{tab:results_quantitative}
\end{table}
\normalsize

%% file: discussion.tex
\section{Conclusion}\label{sec:timelapse_discussion}
In this work, we introduce a new video synthesis problem: making time lapse videos that depict the creation of paintings. We proposed a recurrent probabilistic model that captures the stochastic decisions of human artists. We introduced an alternating sequential training scheme that encourages the model to make realistic predictions over many time steps. We demonstrated our model on digital and watercolor paintings, and used it to synthesize realistic and varied painting videos. Our results, including human evaluations, indicate that the proposed model is a powerful first tool for capturing stochastic changes from small video datasets. 
\section{Acknowledgments}
We thank Zoya Bylinskii of Adobe Inc. for her insights around designing effective and accurate user studies. This work was funded by Wistron Corporation.

%% file: supplementary_derivation.tex
\section{ELBO derivation}
We provide the full derivation of our model and losses from Equation \eqref{eq:optimization}. We start with our goal of finding model parameters $\theta$ that maximize the following probability for all videos and all $t$:
\begin{align*}
&p_\theta(\delta_t,x_{t-1}; x_T)\nonumber\\\vspace{-5pt}
\propto &\hspace{1pt}p_\theta(\delta_t|x_{t-1}; x_T)\nonumber\\\vspace{-5pt}
= &\int_{z_t}\hspace{-2pt}p_\theta(\delta_t|z_t,x_{t-1};x_T)p(z_t)dz_t.
\end{align*}

We use variational inference and introduce an approximate posterior distribution $q_\phi(z_t|\delta_t,x_{t-1};x_T)$ \cite{kingma2014semi,xue2016visual,yan2016attribute2image}.
\small
\begin{align}
&\int_{z_t}p_\theta(\delta_t|z_t,x_{t-1};x_T)p(z_t)dz_t\nonumber\\
=&\int_{z_t}p_\theta(\delta_t|z_t,x_{t-1};x_T)p(z_t)\frac{q_\phi(z_t|\delta_t,x_{t-1};x_T)}{q_\phi(z_t|\delta_t,x_{t-1};x_T)}dz_t\nonumber\\
\propto&\log\int_{z_t}p_\theta(\delta_t|z_t,x_{t-1};x_T)p(z_t)\frac{q_\phi(z_t|\delta_t,x_{t-1};x_T)}{q_\phi(z_t|\delta_t,x_{t-1};x_T)}dz_t\nonumber\\
=&\log\int_{z_t}\frac{p_\theta(\delta_t|z_t,x_{t-1};x_T)p(z_t)}{q_\phi(z_t|\delta_t,x_{t-1};x_T)}q_\phi(z_t|\delta_t,x_{t-1};x_T)dz_t\nonumber\\
=&\log\mathbb{E}_{z\sim q_\phi(z_t|\delta_t,x_{t-1};x_T)}\bigg[\frac{p_\theta(\delta_t|z_t,x_{t-1};x_T)p(z_t)}{q_\phi(z_t|\delta_t,x_{t-1};x_T)}\bigg].
\end{align}
\normalsize
We use the shorthand $z_t\sim q_\phi$ for $z\sim q_\phi(z_t|\delta_t,x_{t-1};x_T)$, and apply Jensen's inequality:
\begin{align}
&\log\mathbb{E}_{z_t\sim q_\phi}\bigg[\frac{p_\theta(\delta_t|z_t,x_{t-1};x_T)p(z_t)}{q_\phi(z_t|\delta_t,x_{t-1};x_T)}\bigg]\nonumber\\
\geq&\hspace{1pt}\mathbb{E}_{z_t\sim q_\phi}\big[\log p_\theta(\delta_t|z_t,x_{t-1};x_T)\big]\nonumber\\
&+ \mathbb{E}_{z\sim q_\phi}\big[\log\frac{p(z_t)}{q_\phi(z_t|\delta_t,x_{t-1};x_T)}\big]\nonumber\\
\geq&\hspace{1pt}\mathbb{E}_{z_t\sim q_\phi}\big[\log p_\theta(\delta_t|z_t,x_{t-1};x_T)\big]\nonumber\\
&- KL[q_\phi(z_t|\delta_t,x_{t-1};x_T)||p(z_t)],
\end{align}\label{eq:elbo_app}
\normalsize
where $KL[\cdot||\cdot]$ is the Kullback-Liebler divergence, arriving at the ELBO presented in Equation \eqref{eq:elbo} in the paper. 

Combining the first term in Equation \eqref{eq:elbo} with our image likelihood defined in Equation \eqref{eq:prob_form_likelihood}:
\begin{align}
&\mathbb{E}_{z_t\sim q_\phi} \log p_\theta(\delta_t|z_t,x_{t-1};x_T)\nonumber\\
\propto&\hspace{1pt}\mathbb{E}_{z_t\sim q_\phi}\big[ \log e^{-\frac{1}{\sigma_1}|\delta_t - \hat{\delta}_t|}\nonumber\\
&+\log\mathcal{N}\big(V(x_{t-1} + \delta_t); V(x_{t-1} + \hat{\delta}_t), \sigma_2^2\mathbbm{I}\big)\big]\nonumber\\
=&\mathbb{E}_{z_t\sim q_\phi}\bigg[{-\frac{1}{\sigma_1}|\delta_t - \hat{\delta}_t|}\nonumber\\
&+\log\frac{1}{\sqrt{2\pi \sigma_2^2}}\exp\big(-\frac{(V(x_{t-1} + \delta_t)- V(x_{t-1} + \hat{\delta}_t))^2}{2\sigma_2^2}\big)\bigg]\nonumber\\
\propto&\hspace{1pt}\mathbb{E}_{z_t\sim q_\phi}\bigg[{-\frac{1}{\sigma_1}|\delta_t - \hat{\delta}_t|}\nonumber\\
&-\frac{1}{2\sigma_2^2}(V(x_{t-1} + \delta_t)- V(x_{t-1} + \hat{\delta}_t))^2\bigg], 
\end{align}
giving us the image similarity losses in Equation \eqref{eq:loss_pairwise}. We derive $\mathcal{L}_{KL}$ in Equation \eqref{eq:loss_pairwise} by similarly taking the logarithm of the normal distributions defined in Equations \eqref{eq:prob_form_prior} and \eqref{eq:prob_form_posterior}.

%% file: supplementary_details.tex
\section{Network architecture}
\begin{figure*}
\includegraphics[width=\linewidth]{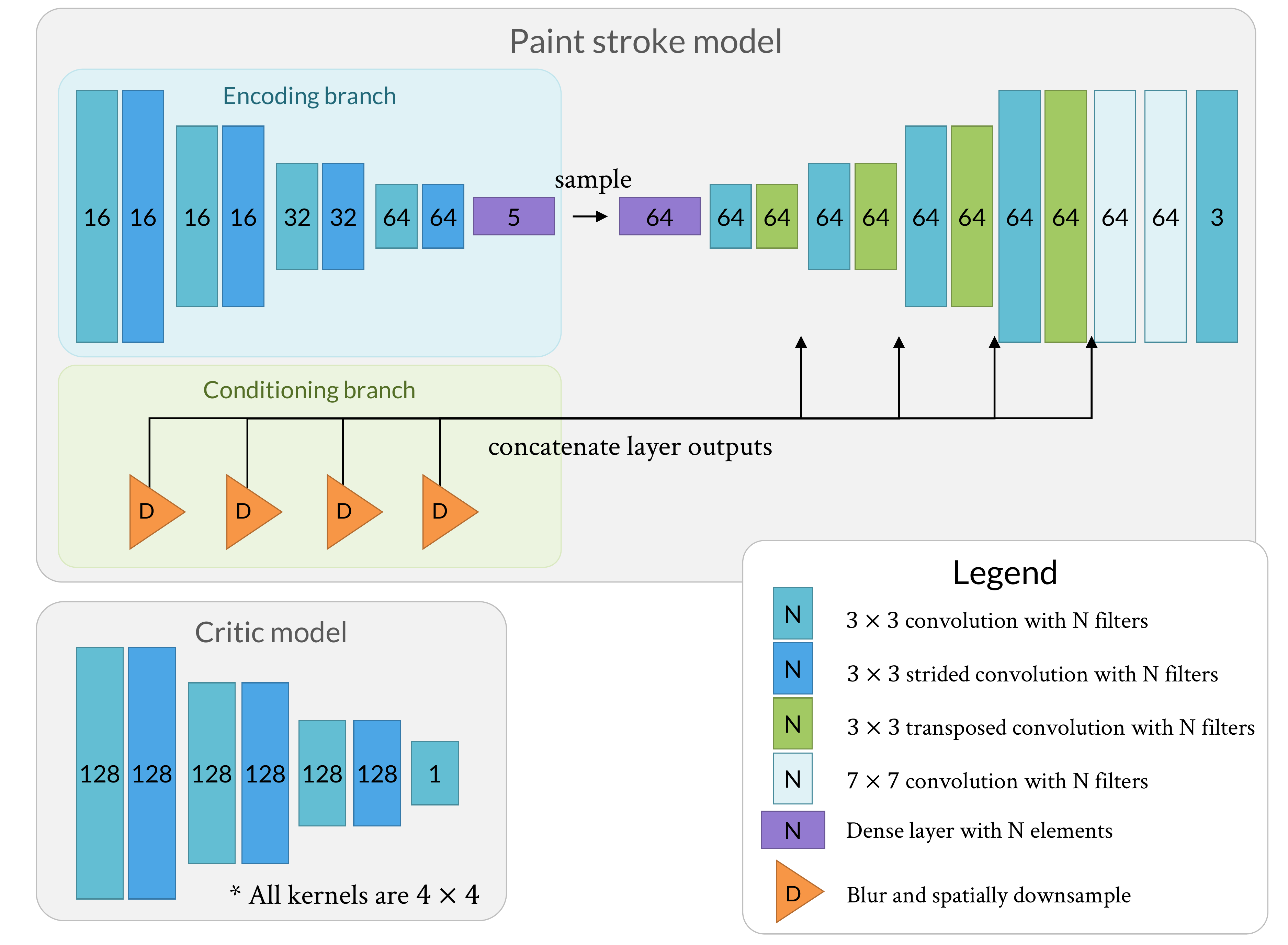}
\caption{\textbf{Neural network architecture details}. We use an encoder-decoder style architecture for our model. For our critic, we use a similar architecture to StarGAN \cite{choi2018stargan}, and optimize the critic using WGAN-GP \cite{gulrajani2017improved} with a gradient penalty weight of 10, and 5 critic training iterations for each iteration of our model. All strided convolutions and downsampling layers reduce the size of the input volume by a factor of 2. } \label{fig:supplementary_networks}
\end{figure*}

We provide details about the architecture of our recurrent model and our critic model in Figure \ref{fig:supplementary_networks}.
\section{Human study}
We surveyed 150 human participants. Each participant took a survey containing a training section followed by 14 questions. 

\begin{enumerate}[leftmargin=0pt,topsep=2pt,itemsep=0pt]
\item[] \textbf{Calibration}: We first trained the participants by showing them several examples of real digital and watercolor painting time lapses. 

\item[] \textbf{Evaluation}:
We then showed each participant 14 pairs of time lapse videos, comprised of a mix of watercolor and digital paintings selected randomly from the test sets. Although each participant only saw a subset of the test paintings, every test painting was included in the surveys. Each pair contained videos of the same center-cropped painting. The videos were randomly chosen from all pairwise comparisons between real, \textit{vdp}, and \textit{ours}, with the ordering within each pair randomized as well. Samples from \textit{vdp} and \textit{ours} were generated randomly.

\item[] \textbf{Validation}: Within the survey, we also showed two repeated questions comparing a real video with a linearly interpolated video (which we described as \textit{interp} in Table \ref{tab:results_quantitative} in the paper) to validate that users understood the task. We did not use results from users who chose incorrect answers for one or both validation questions. 
\end{enumerate}

%% file: supplementary_qualitative.tex
\section{Additional results}
\begin{figure*}[t]
\centering
\subfloat[\textbf{The proposed method paints similar regions to the artist}. Red arrows in the second row show where \textit{unet} adds fine details everywhere in the scene, ignoring the semantic boundary between the rock and the water, and contributing to an unrealistic fading effect. The video synthesized by \textit{vdp} \hl{produces} more coarse \hl{changes} early on, but introduces an unrealistic-looking blurring and fading effect on the rock (red arrows in the third row). Blue arrows highlight that our method makes similar changes to the artist, filling in the base color of the water, then the base colors of the rock, and then fine details throughout the painting.]{
\includegraphics[width=\linewidth]{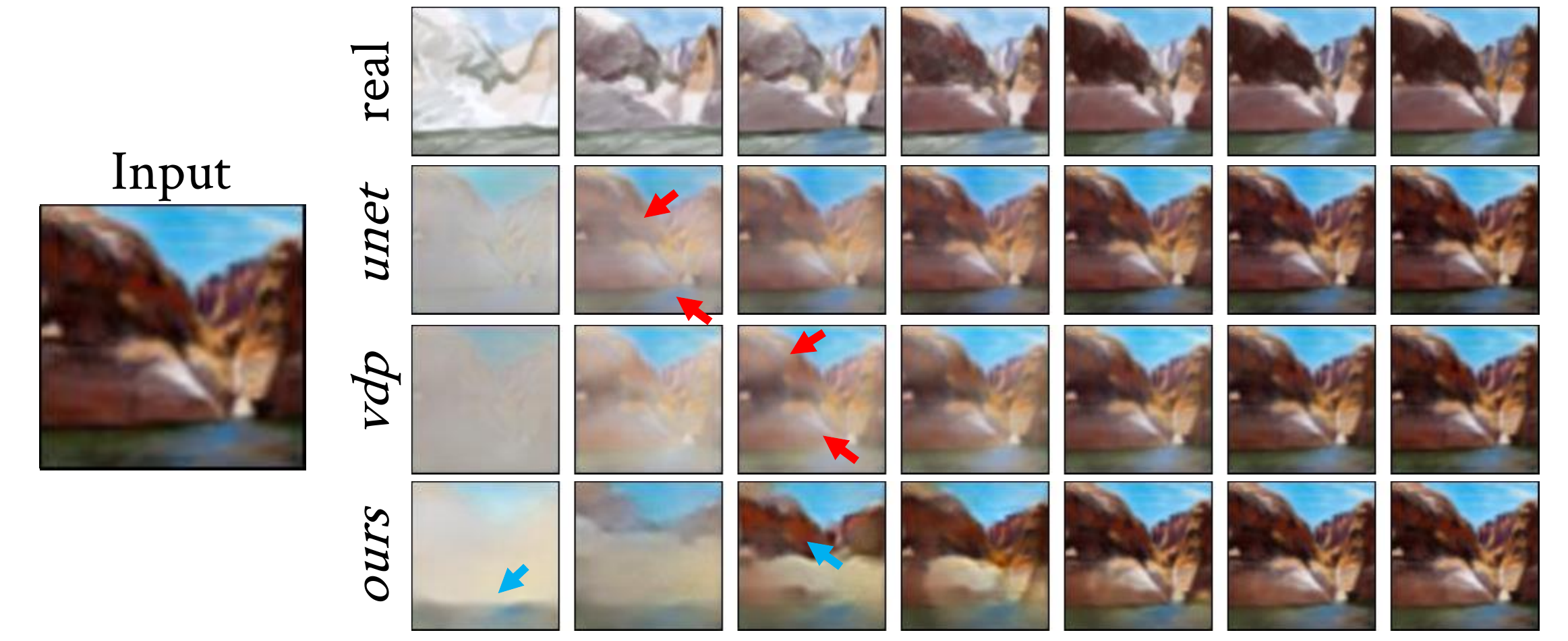}
}\\
\subfloat[\textbf{The proposed method identifies appropriate colors and shape for each layer of paint}. Red arrows indicate where the baselines fill in details that the artist does not complete until much later in the sequence (not shown in the real sequence, but visible in the input image). Blue arrows show where our method adds a base layer for the vase with a reasonable color and shape, and then adds fine details to it later.]{
\includegraphics[width=\linewidth]{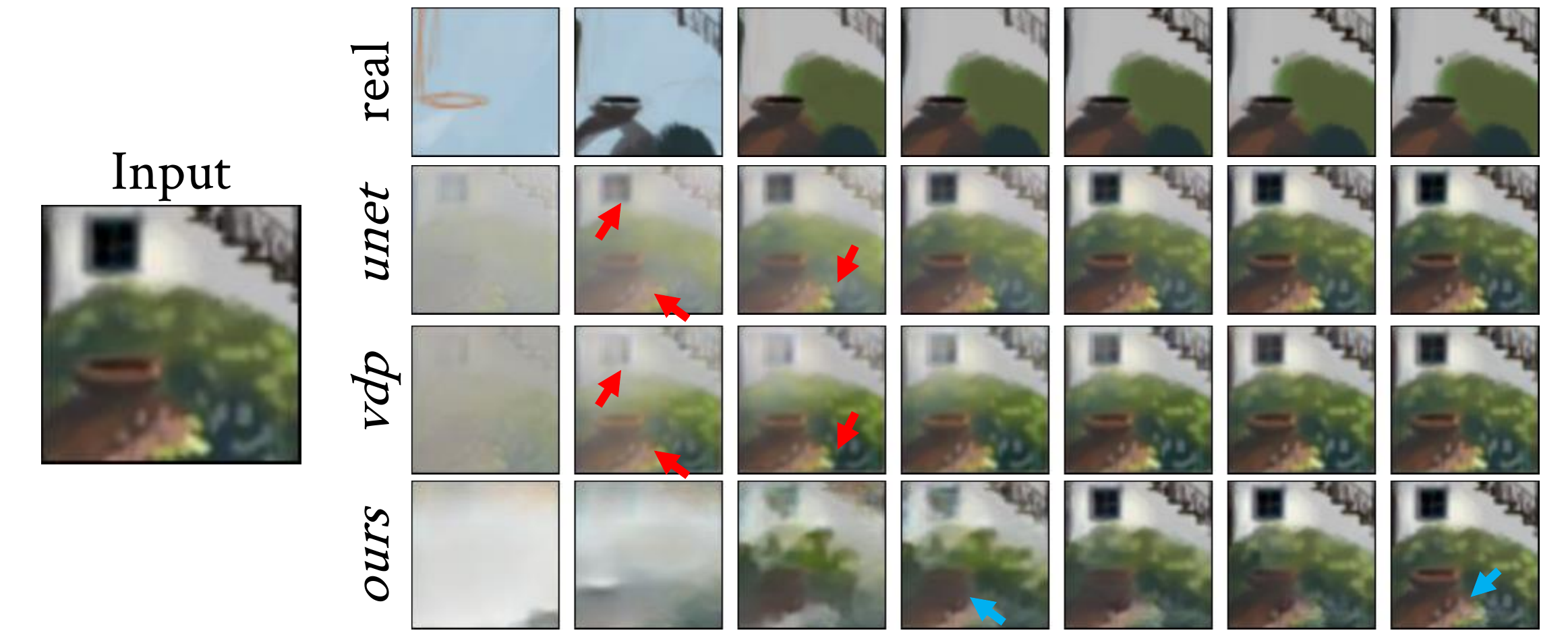}
}\caption{\textbf{Videos synthesized from the watercolor paintings test set}. For the stochastic methods \textit{vdp} and \textit{ours}, we examine the nearest sample to the real video out of 2000 samples. We discuss the variability among samples from our method in Section \ref{sec:timelapse_experiments}, and in the supplementary video.}\label{fig:supp_results_digital}
\end{figure*}

\begin{figure*}[t]
\centering
\subfloat[\textbf{The proposed method paints using coarse-to-fine layers of different colors, similarly to the real artist}. Red arrows indicate where the baseline methods fill in details of the house and bush at the same time, adding fine-grained details even early in the painting. Blue arrows highlight where our method makes similar changes to the artist, adding a flat base color for the bush first before filling in details, and using layers of different colors.]{
\includegraphics[width=\linewidth]{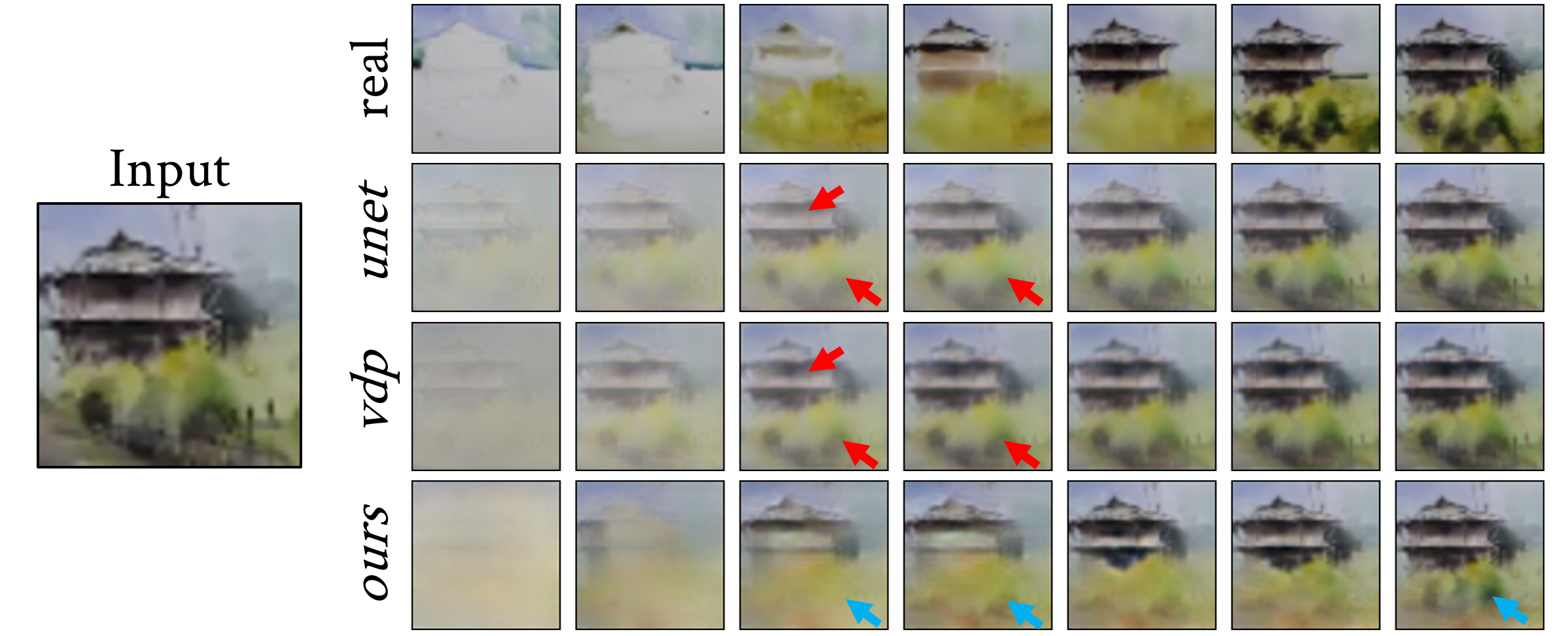}
}\label{fig:supp_results_wc_1}
\subfloat[\textbf{The proposed method synthesizes watercolor-like effects such as paint fading as it dries}. Red arrows indicate where the baselines fill in the house and the background at the same time. Blue arrows in the first two video frames of the last row show that our method uses coarse \hl{changes} early on. Blue arrows in frames 3-5 show where our method simulates paint drying effects (with the intensity of the color fading over time), which are common in real watercolor videos.]{
\includegraphics[width=\linewidth]{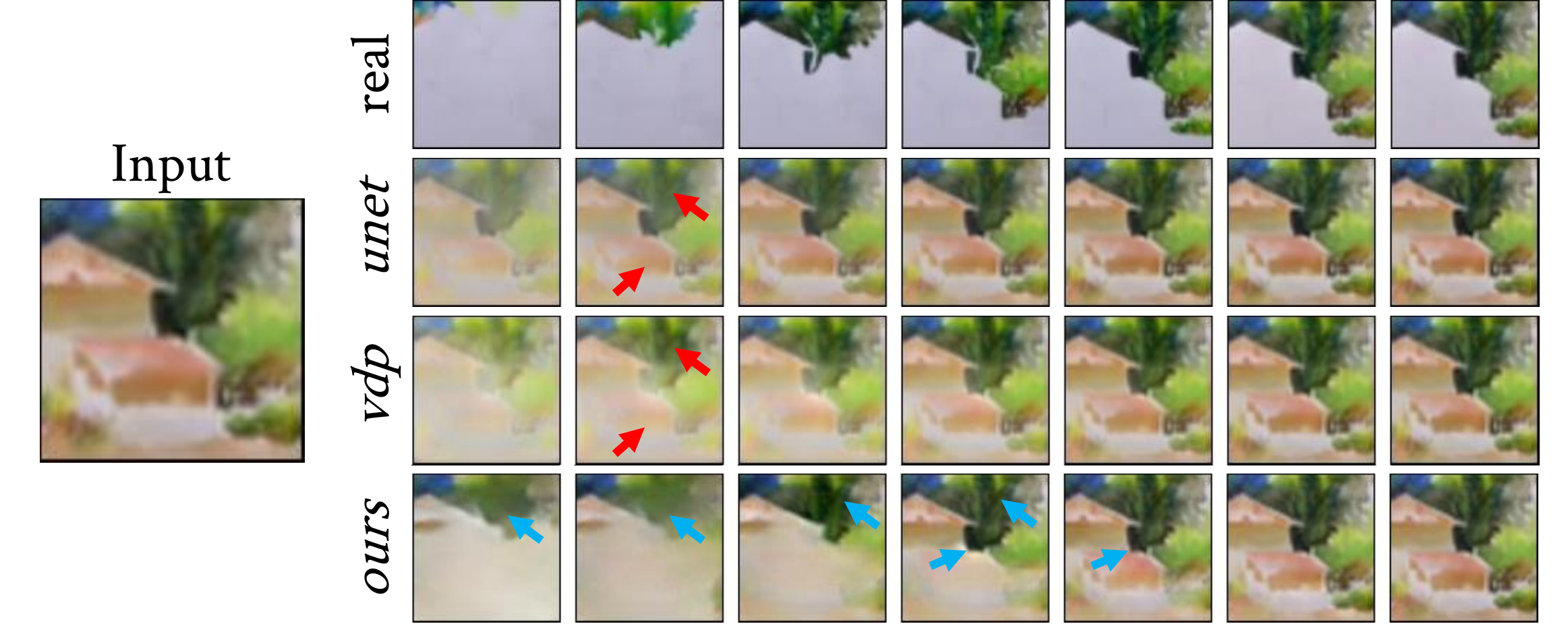}
}\caption{\textbf{Videos synthesized from the watercolor paintings test set}. For the stochastic methods \textit{vdp} and \textit{ours}, we show the nearest sample to the real video out of 2000 samples. }\label{fig:supp_results_watercolor}
\end{figure*}

We include additional qualitative results in Figures \ref{fig:supp_results_digital} and \ref{fig:supp_results_watercolor}. We encourage the reader to view the supplementary video, which illustrates many of the discussed effects. 

